\documentstyle[11pt,moriond,epsfig]{article}
\def\Journal#1#2#3#4{{#1} {\bf #2}, #3 (#4)}

\def\PLB{{\em Phys. Lett.}  B}
\def\PRL{\em Phys. Rev. Lett.}
\def\PRD{{\em Phys. Rev.} D}

\def\be{\begin{equation}}
\def\ee{\end{equation}}
\def\bea{\begin{eqnarray}}
\def\eea{\end{eqnarray}}

\begin{document}
\vspace*{4cm}
\title{MODELLING BOSE-EINSTEIN EFFECT FROM ASYMMETRIC SOURCES
IN MONTE CARLO GENERATORS\footnote{Presented by K. Fia{\l}kowski at the 
XXXVth Rencontres de MORIOND, "QCD and High Energy Hadronic Interactions", 
to be published in the Proceedings.}}

\author{K. FIA{\L}KOWSKI and R. WIT}

\address{Institute of Physics, Jagellonian University,\\
Reymonta 4, 30-059 Krakow, Poland}

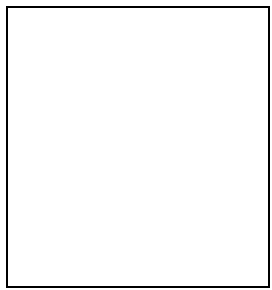
\maketitle
\abstracts{
We present some results  concerning the Bose-Einstein effect in asymmetric
sources in  Monte Carlo generators. A comparison of LEP  data,
standard JETSET  predictions and results from the weight method
is given. Possible generalization of the weight factors to an asymmetric
form is considered. }
\pagebreak

\section{Introductory remarks and data}
Recently we observe a renewal of interest in analysing
the space-time structure of sources in multiparticle production by means
of Bose-Einstein (BE) interference. Such analysis followed the example of
astrophysical investigations of Hanbury-Brown and Twiss~\cite{HBT}. By
 improving the standard approach~\cite{BGJ} it became possible to model
this effect in Monte Carlo generators: as the "afterburner" for which the
original MC provides a source~\cite{SUL,zha}, by shifting the
momenta~\cite{SJO} or by adding weights to generated events~\cite{BK,FWW}.
\par
The analysis of BE effect in 3 dimensions is supposed to reflect the
spatial source asymmetry.  Such analysis  was done for the LEP data at the
$Z^0$ peak~\cite{CU} which have very high statistics and good accuracy.
In the following we concentrate our attention
on the L3 data~\cite{L3}, as the DELPHI data~\cite{DEL} are parametrized
with only two radii, and the OPAL data~\cite{OPAL} use the like/unlike
ratio which requires a cut off of the resonance affected regions even in
double ratios.
\par
As in the L3 paper~\cite{L3} we use for each pair of identical pions
three components of the invariant
$Q^2={-(p_1-p_2)^2}$: $Q_L^2, Q_{out}^2, Q_{side}^2$ defined in the LCMS,
where the sum of three - vector momenta is
perpendicular to the thrust axis. Similarly we  define a "double ratio"
using a  reference sample from mixed events:  $$ R_2(p_1, p_2) =
\frac{\rho_2}{\rho_2^{mix}} / \frac{\rho_2^{MC}}{\rho^{mix, MC}_{2}}.$$
 This "double ratio" is
parametrized by $$ R_2(Q_L,Q_{out},Q_{side}) = \gamma [1+ \delta Q_L+
 \epsilon Q_{out}+ \zeta Q_{side}]~ \cdot ~~~~~~~~~~~~~~~~~~~~ $$
$$
~~~~~~~~~~\cdot ~ [1+ \lambda exp(-R^2_L Q_L^2-R^2_{out}Q^2_{out}-
R^2_{side}Q^2_{side}-
2\rho_{L,out}R_{L}R_{out} Q_{L} Q_{out})]
$$
The first bracket reflects possible traces of
long-distance correlations; the last term in the second bracket seems to
be negligible when fitting data and will be omitted in the following.
\par
By fitting the parameters $R_L$ and $R_{side}$ we get some information on
the geometric radii in the longitudinal and transverse directions
(respective to the thrust axis). $R_{out}$ reflects both the spatial
extension and time duration of the emission process.
\par
In the L3 data the fit region in all three variables extends to
1.04 GeV and is divided into 13 bins, which gives 2197 points fitted with
8 parameters.  The fit parameters $\delta, \epsilon$ and $\zeta$ are
rather small; this means the observed BE enhancement is rather well
approximated with a Gaussian.  \par The fitted values of radii (in {\bf
fm}) are as follows:  $$ R_L = 0.74 \pm 0.02^{+0.04}_{-0.03},~ R_{out} =
0.53 \pm 0.02^{+0.05}_{-0.06},~ R_{side} = 0.59 \pm 0.01^{+0.03}_{-0.13}$$
We see clear evidence for source elongation:
$R_{side}/R_L$ is smaller than one by more than four standard deviations.
\section{Asymmetric effects from symmetric models}
The geometric interpretation of data requires  a comparison with the
results from the
standard MC procedures modelling the BE effect.  In the L3 paper such
an analysis is given for the standard LUBOEI procedure built into the
JETSET Monte Carlo generator.This procedure modifies the final state by a
shift of momenta for each pair of identical pions. The shift is calculated
to enhance low values of $Q^2$ and to reproduce the experimental ratio in
this variable. The superposition of the procedure for all the pairs and
subsequent rescaling (restoring the energy conservation) makes the
connection between the parameters of the shift and the resulting ratio
rather indirect.  \par Using the JETSET parameters adjusted to all the L3
data and the LUBOEI parameters fitted to describe the BE ratio in $Q^2$
the authors of the L3 paper calculated the same quantities as measured in
the experiment.  The projections of $R_2$ look qualitatively very similar
to the experimental ones. However, the fit to the 3-dimensional
distribution gives results different from data. The ratio $R_{side}/R_L$
is not smaller but greater than one; the fitted values (in {\bf fm}) are:
$$R_L= 0.71 \pm 0.01, R_{out} = 0.58 \pm 0.01, R_{side} = 0.75 \pm 0.01.$$
We confirmed these numbers in our calculations. We found also
that the results are  sensitive to the JETSET parameters. Using the
default values instead of the L3 values we obtained
a significantly smaller value of $R_{out}$ (below 0.5) and significantly
smaller $\lambda$. Other values are less affected and $R_{side}/R_l$ still
bigger than 1.
\par
\begin{figure}[h]
%\vspace{11cm}
\centerline{
\epsfig{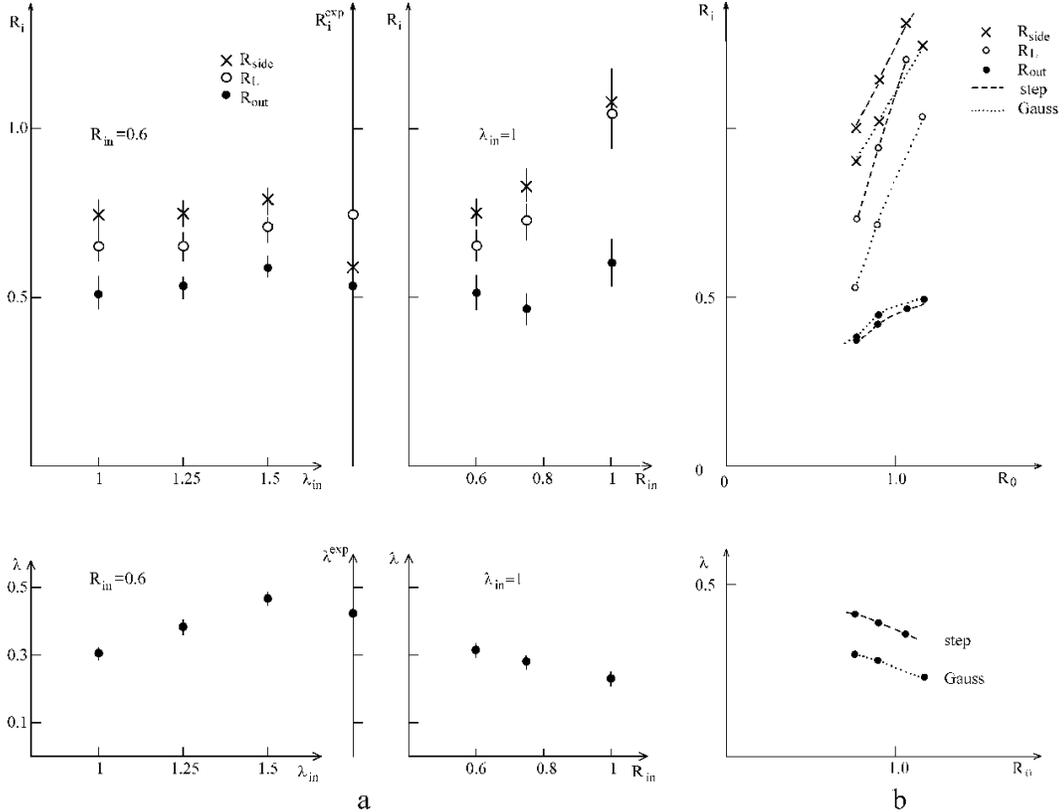}}
\caption{Fit parameters $\lambda$ and $R_i$ as functions of the
input parameters a) for the LUBOEI procedure, b) for the weight method.
Experimental values are shown on a separate vertical axis.}
\label{fig:fig1}
\end{figure}

We have checked also how the results depend on the source radius $R$ and
incoherence parameter $\lambda_{in}$ assumed in the LUBOEI input function
$R_{BE}(Q)=1+\lambda_{in} exp(-R^2 Q^2)$.  In all cases we get
$R_{side}>R_L>R{out}$, although the input function was obviously
symmetric.  The values of $R_{side}$ and $R_L$ are proportional to $R$,
whereas $R_{out}$ changes much less; the dependence on $\lambda_{in}$ is
very weak.  The output value of $\lambda$ decreases quite strongly with
increasing $R$ and increases with $\lambda_{in}$. No choice of input
parameters gives the values of $R_i$ compatible with data.  This is shown
in Fig.~\ref{fig:fig1}.
\par Another interesting observation is that to
fit the L3 data one needs $\lambda = 1.5$, which is beyond the physically
acceptable value of 1. This supports our doubts about the usefulness of
the LUBOEI procedure in understanding the experimental results (although
certainly it is the most practical description of data).
\par Therefore
we have also compared the data with the results from another procedure
modelling the BE effect - the weight method~\cite{FWW}. In this method
each event gets a weight calculated by summing the products of two
particle weight factors, which are just 1 for equal momenta and vanish for
large separation in momentum space. A reasonable description of the effect
in $Q^2$ is obtained with a simple gaussian form of the weight factor
\begin{equation}
w_2(p,q)=exp[-(p-q)^2R_0^2/2],
%\label{eq:wf}
\end{equation}
or, even simpler, $\theta $ - function form with $w_2 = 1 $ for some
range of $-(p-q)^2<1/R_0^2$ and $w_2=0$ outside.
\par
In this method we may repeat the same calculation as done for the LUBOEI
procedure. The resulting double ratios are not that smooth and monotically
decreasing as in the data or from the LUBOEI procedure (which is the
usual drawback of the weight methods).  However, the major features are
surprisingly similar:  with weight factors depending only on $Q^2$ we get
different values of fitted $R_i$ parameters.  Moreover, the hierarchy of
parameters is the same:  $R_{side} > R_L$. This suggests that the
assymetry is generated by the jet-like structure of final states
and not by any specific features of the procedure modelling the BE effect.
In Fig. 1b we show the values of the fit parameters as functions of $R_0$
for a Gaussian as well as  the $\theta$-like weight factors.
Again, no choice of input parameters allows to describe the data.
\par
These results suggest also that one should be careful
with the geometric interpretation of the data. If one gets asymmetric
distributions from the generator without assuming explicitly space
asymmetry of the source, it is not clear how the assumed asymmetry will be
reflected in the results.
\par
We tried to get some information on this
problem within the asymmetric weight method, i.e.  introducing weight
factors which depend in a different way on $Q_L$, $Q_{side}$ and
$Q_{out}$.  This work is progress.  However, it seems rather difficult to
reproduce the data even with two more parameters.

\section{Conclusions and outlook}
In this note we investigated the asymmetry of the BE effect in two
procedures imitating this effect in the Monte Carlo generators and
compared them to the data  at $Z^0$ peak. Both procedures give
surprisingly similar results and disagree with data. The work on the
possible introduction of asymmetry in the weight method is in progress.
\section*{Acknowledgments}
This work was partially supported by the KBN grants No 2 P03B 086 14, 2
P03B 010 15 and  2 P03B 019 17.


\section*{References}
\begin{thebibliography}{99}
\bibitem{HBT}R. Hanbury-Brown and R.Q. Twiss, \Journal {\it Nature}{178}
{1046}{1956}.
\bibitem{BGJ}D.H. Boal, C.-K. Gelbke and B.K. Jennings, \Journal
{\it Rev. Mod. Phys.}{62}{553}{1990}.
\bibitem{SUL}J.P. Sullivan {\it et al.}, \Journal{\PRL}{70}{3000}{1993}.
\bibitem{zha}Q.H. Zhang {\it et al.}, \Journal{\PLB}{407}{33}{1997}.
\bibitem{SJO}T. Sj\"ostrand and M. Bengtsson, \Journal{\it Comp. Phys.
 Comm.}{43}{367}{1987}; T. Sj\"ostrand, \Journal{\it Comp. Phys. Comm.}
{82}{74}{1994}.
\bibitem{BK}A. Bia{\l}as and A. Krzywicki, \Journal{\PLB}{354}{134}{1995}.
\bibitem{FWW}K. Fia{\l}kowski, R. Wit and J. Wosiek,
\Journal{\PRD}{57}{094013}{1998}.
\bibitem{CU}M. Cuffiani, talk at this meeting.
\bibitem{L3} The L3 Collaboration, \Journal{\PLB}{458}{517}{1999}.
\bibitem{DEL} The DELPHI Collaboration, \Journal{\PLB}{471}{460}{2000}.
\bibitem{OPAL} The OPAL Collaboration, CERN preprint CERN-EP-2000-004.
\end{thebibliography}
\end{document}